\pdfoutput=1
\documentclass[%
 reprint,
superscriptaddress,
 amsmath,amssymb,
 aps,
 pra,
]{revtex4-2}

\usepackage{caption}
\usepackage{subcaption}
\usepackage{graphicx}
\usepackage{amsmath}
\usepackage{geometry}
\usepackage{hyperref}
\usepackage[T1]{fontenc}
\usepackage[utf8]{inputenc}
\usepackage{float}
\usepackage{csquotes}
\usepackage[version=4]{mhchem}
\usepackage{amssymb}
\usepackage{xcolor}
\usepackage{bbold}
\usepackage{dcolumn}
\usepackage{bm}
\usepackage{algorithm}
\usepackage{algpseudocode}

\begin{document}
\preprint{}

\title{
Knowledge Distillation Inspired Variational Quantum Eigensolver with Virtual Annealing
}

\author{Junxu Li}
\email{Emial: lijunxu1996@gmail.com}
\affiliation{Department of Physics, College of Science, Northeastern University, Shenyang 110819, China}
\date{\today}

\begin{abstract}
In this paper, we propose a Knowledge Distillation Inspired Variational Quantum Eigensolver (KD-VQE).
Inspired by the virtual distillation process in knowledge distillation (KD), KD-VQE introduces a virtual annealing mechanism to the variational quantum eigensolver (VQE) framework.
In KD-VQE, measurement resources (shots) are dynamically allocated among multiple trial wavefunctions, each weighted according to a Boltzmann distribution with a virtual temperature.
As the temperature decreases gradually, the algorithm progressively reallocates resources toward lower-energy candidates, effectively filtering out suboptimal states and steering the system toward the global minimum.
Moreover, we demonstrate the effectiveness of KD-VQE by applying it to the two-site Fermi-Hubbard model.
Compared to standard VQE framework, KD-VQE explores a broader region of the solution space, and offers improved convergence behavior and increased reliability.
\end{abstract}

\maketitle

\section{Introduction}
The variational quantum eigensolver (VQE)\cite{peruzzo2014variational, mcclean2016theory}, firstly introduced by Peruzzo {\it et al.}, is one of the most promising hybrid quantum-classical algorithms for addressing problems that are intractable on classical computers\cite{preskill2018quantum, bittel2021training}.
VQE leverages the variational principle to estimate the ground state energy of a given Hamiltonian.
The scope of VQE is broad, encompassing applications in fields such as quantum chemistry\cite{o2016scalable, hempel2018quantum, cao2019quantum, parrish2019quantum, google2020hartree, cerezo2021variational, tilly2022variational, sajjan2022quantum, ma2023multiscale, guo2024experimental,kim2024qudit, delgado2025quantum}, condensed matter physics\cite{kandala2017hardware, kokail2019self, uvarov2020variational, mizuta2021deep, gyawali2022adaptive, anselme2022simulating, kattemolle2022variational}, and beyond.
In essence, VQE is an optimization algorithm that heuristically constructs an approximate wavefunction through the iterative adjustment of ansatz parameters.
For such variational methods, it is critical that they can converge to high-quality approximations within a tractable number of optimization steps and measurements\cite{wecker2015progress, mcclean2018barren, wierichs2020avoiding}.
Hence, the design of efficient optimization strategies is of great importance for ensuring the scalability and reliability of VQE.

To improve the optimization process, we propose a knowledge distillation inspired variational quantum eigensolver (KD-VQE).
This approach draws inspiration from knowledge distillation (KD)\cite{hinton2015distilling}, a powerful technique widely used to optimize deep neural networks\cite{chen2017learning, cho2019efficacy, cheng2020explaining, ji2020knowledge, gu2023minillm, liu2024deepseek}.
Despite the remarkable success of deep neural networks, deploying large, complex models on devices with limited computational resources, such as the mobile phones and embedded devices, remains a considerable challenge\cite{gou2021knowledge}.
The cumbersome models "assign probabilities to all of the incorrect answers and even when these probabilities are very small, some of them are much larger than others"\cite{hinton2015distilling}.
KD is proposed to transfer the generalization ability of the cumbersome `teacher' model to a small `student' model, which is better suited for deployment.
Typically, KD is implemented through a virtual `distillation' process.
This involves replacing the softmax output layer of the teacher network with a layer of sigmoid functions, where the parameters are controlled by a virtual Boltzmann temperature\cite{hinton2015distilling, gou2021knowledge}.

A similar challenge arises in standard VQE, where the optimization process is sensitive to the choice of the initial trial wavefunction\cite{jattana2023improved, skogh2023accelerating}.
If the initial guess is close to the true ground state, convergence is efficient.
On the contrary, poor initialization, such as proximity to excited states or local minima, can lead to significant overhead in both time and resources.
However, the standard VQE framework can not distinguish the good and poor candidates.
The initial guesses are treated in the same manner, optimized until they converge to the global or a local minima.
Like the cumbersome teacher model in KD, standard VQE spends much computational resources on the suboptimal candidates.

To address this issue, we introduce a virtual annealing process in KD-VQE, which is inspired by the virtual distillation process in KD.
KD-VQE utilizes a collection of trial wavefunctions as a mixed-state, which is governed by a Boltzmann distribution with a virtual temperature.
Unlike standard VQE, which focuses on a single variational state, KD-VQE dynamically allocates measurement resources among multiple candidates.
As the virtual temperature gradually decreases from infinity to zero, the ensemble of wavefunctions progressively condenses toward the lowest-energy state. 
This virtual annealing process allows KD-VQE to filter out suboptimal states and prioritize the most promising solutions.
Compared to standard VQE framework, KD-VQE explores a broader region of the solution space, and offers improved convergence behavior and increased reliability.
In the rest of this paper, we will demonstrate the detailed pipeline of KD-VQE, along with a simple application solving 2-site Fermi Hubbard model.

\section{Pipeline of KD-VQE}
\begin{figure}[t]
    \begin{center}
        \includegraphics[width=0.48\textwidth]{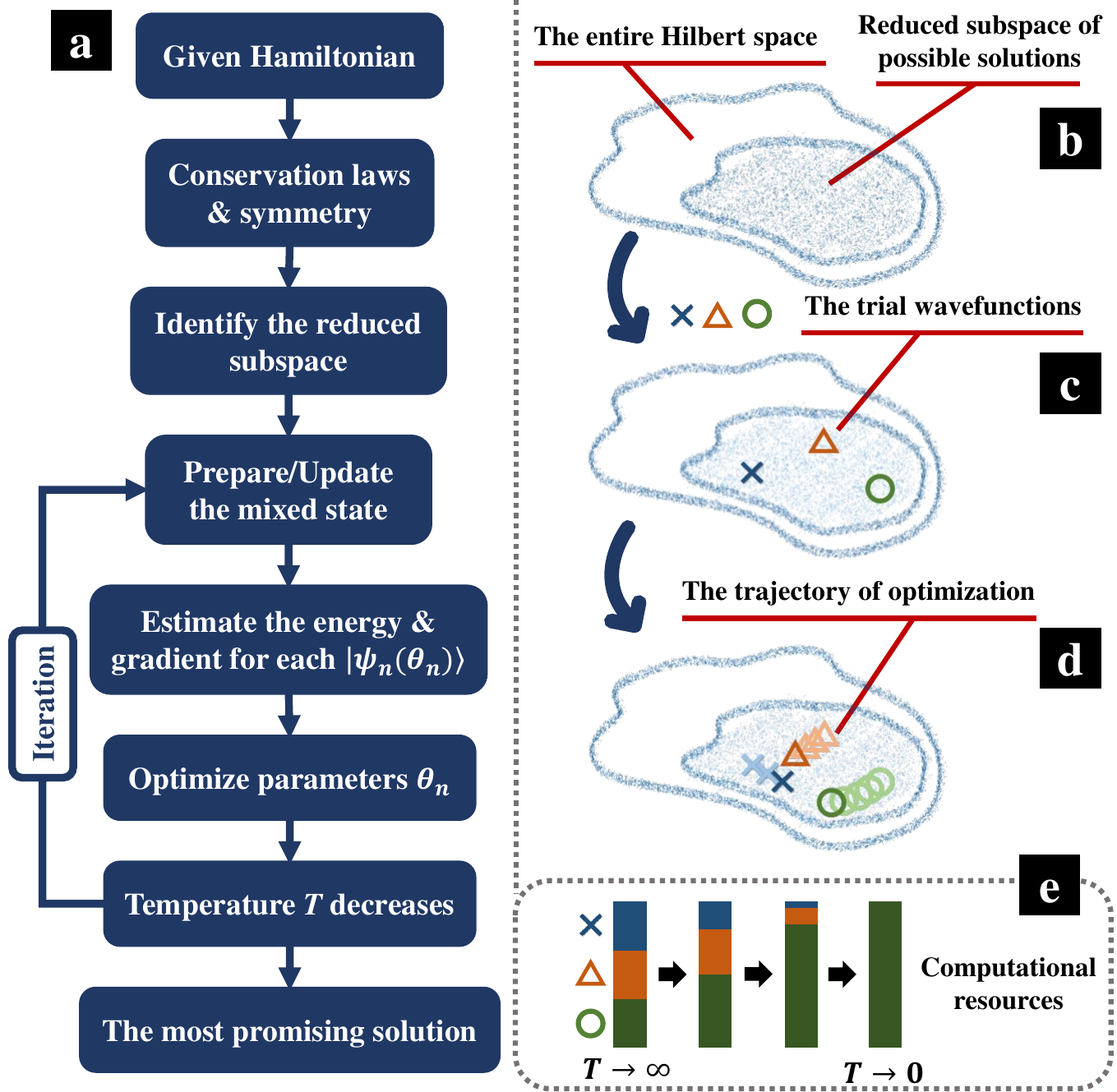}
    \end{center}
    \caption{
    {\bf Schematic diagrams of the KD-VQE pipeline.}
    (a) Overview of the KD-VQE pipeline.
    (b-d) Schematic illustrations of the key stages in the KD-VQE process.
    (e) Computational resources are dynamically reallocated to the most promising wavefunction.
    }
    \label{fig_pipeline}
\end{figure}

In Fig.(\ref{fig_pipeline}a) we present the KD-VQE pipeline, which consists three main parts.

1. Identification of the Reduced Subspace.
The first step of KD-VQE is to examine the conservation laws and symmetry properties of the given Hamiltonian $H$.
In KD techniques, pre-trained `teacher' models possess a greater knowledge capacity than the smaller `student' models.
However, in most practical applications, it is unnecessary to fully exploit this capacity.
Similarly, while the Hamiltonian $H$ encapsulates all information about the system, searching the entire Hilbert space for the ground state is neither efficient nor practical.
Therefore, it is essential to identify a reduced subspace of possible solutions by leveraging the conservation laws and symmetry properties of the system.

2. Preparation of the Mixed State Ansatz.
Once the reduced subspace is determined, the next step is to construct the trial wavefunctions and prepare the mixed state according to the Boltzmann distribution.
Initially, several candidate solutions within the reduced subspace are selected, serving as initial guesses for KD-VQE. 
A practical approach is to identify the solvable component of $H$, and use the `eigenstates' of this solvable part as initial guesses.
In the optimization and virtual annealing process, the trial wavefunctions serve as a mixed state.
For simplicity, we denote these trial wavefunctions as $|\psi_k(\theta_k)\rangle$, where $\theta_k$ are variational parameters.
The mixed state is constructed according to the Boltzmann distribution with a virtual temperature $T$,
\begin{equation}
    \rho(T)
    =
    \frac{1}{Z}\sum_k\exp(-\varepsilon_k/T)
    |\psi_k(\theta_k)\rangle\langle \psi_k(\theta_k)|
    \label{eq_rho}
\end{equation}
where $\varepsilon$ represents the estimated energy of the state $|\psi(\theta_k)\rangle$, and the partition function is given by $Z = \sum_k \exp(-\varepsilon_k/T)$.
The implementation of the mixed state is achieved by assigning an appropriate number of measurement shots to each trial wavefunction during the optimization process. 
Given a total number of shots $N_{s}$, each trial wavefunction $|\psi_k(\theta_k)\rangle$ is assigned $N_s\exp(-\varepsilon_k/T)/Z$ shots.

3. Variational Optimization along with Virtual Annealing.
The optimization process in KD-VQE follows the standard VQE method.
For each trial wavefunction $|\psi_k(\theta_k)\rangle$, the variational parameters $\theta_k$ are independently updated by minimizing the expectation value $\langle \psi_k({\theta_k})|H|\psi_k({\theta_k})\rangle$.
The standard VQE framework employs the variational principle to approximate the ground state energy of a given Hamiltonian.
For a parameterized wavefunction  $\psi({\theta})$, the optimization problem of VQE is formulated as
\begin{equation}
    E_{gs}\leq \langle \psi({\theta})|H|\psi({\theta})\rangle
    \label{eq_vqe_opt}
\end{equation}
where $E_{gs}$ indicates the energy of ground state.
The variational parameters $\theta$ are iteratively updated until convergence is achieved.
In the case of mixed states, the variational principle remains applicable.
Consider the mixed state as shown in Eq.(\ref{eq_rho}), Eq.(\ref{eq_vqe_opt}) can be rewritten as
\begin{equation}
    E_{gs}\leq tr(\rho H) = \sum_k 
    \frac{e^{-\varepsilon_k/T}}{Z}
    \langle \psi_k({\theta_k})|H|\psi_k({\theta_k})\rangle
    \label{eq_kdvqe_opt}
\end{equation}
Eq.(\ref{eq_kdvqe_opt}) ensures that KD-VQE can converge to the ground state through the optimization process.
In addition to optimizing the variational parameters, KD-VQE incorporates a virtual annealing mechanism, which dynamically allocates computational resources to the most promising wavefunctions.
In KD-VQE, the virtual temperature $T$ gradually decreases from infinity to zero.
Initially, $T\rightarrow+\infty$, the system corresponds to a uniform mixture, as described by Eq.(\ref{eq_rho}).
As $T$ decreases, higher-energy components are progressively filtered out.
By this mean, the virtual annealing process refines the ansatz and accelerates convergence to the ground state.

\begin{figure*}[t]
    \begin{center}
        \includegraphics[width=0.98\textwidth]{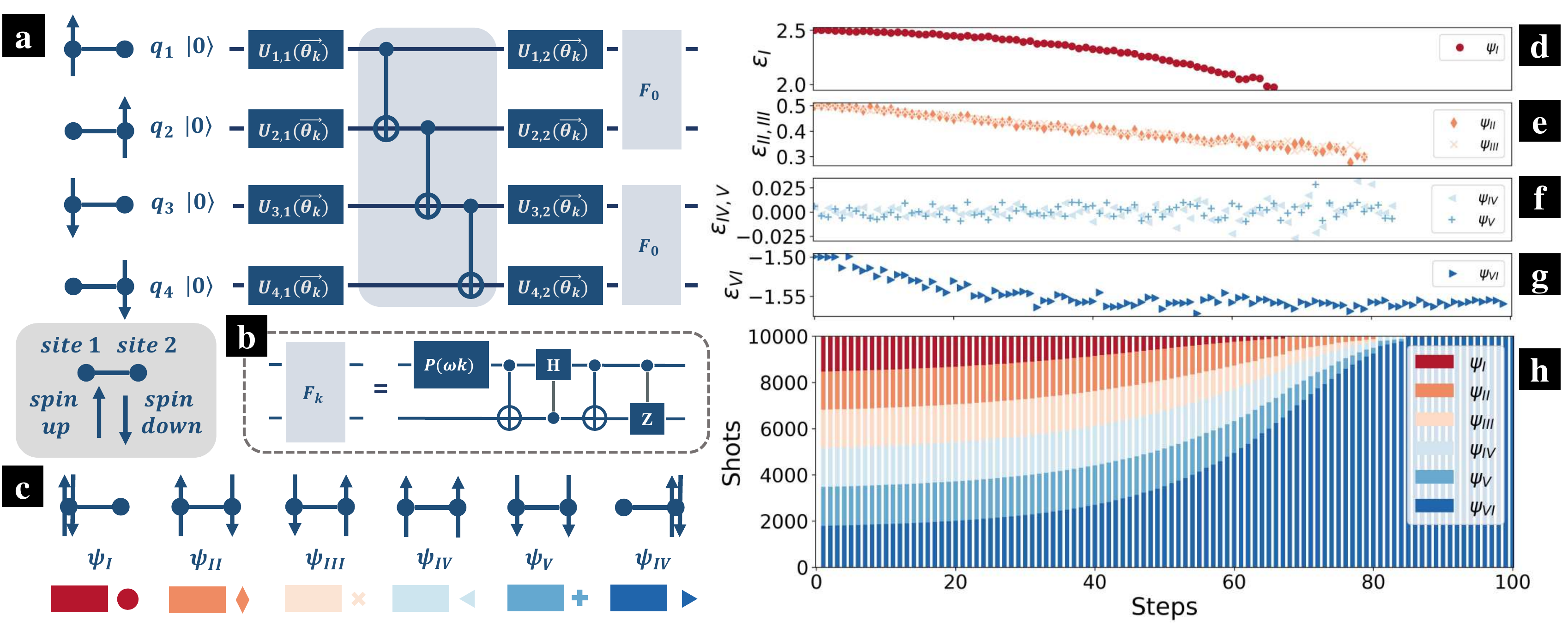}
    \end{center}
    \caption{
    {\bf Schematic diagrams of KD-VQE solving half filling 2-site Fermi-Hubbard model.}
    (a) Ansatz circuit used for preparing trial wavefunctions, where $U(\theta_k)$ parameterized single-qubit gates and the dashed box indicates entangling $CNOT$ gates.
    (b) Quantum circuit implementing the Fourier transformation used to diagonalize the hopping term.
    (c) Six selected half-filling initial states $\psi_{I,\cdots,VI}$, shown before Fourier transformation.
    (d–g) Energy evolution of the trial states during optimization and virtual annealing.
    (h) Corresponding dynamic reallocation of measurement shots, illustrating the gradual concentration of resources toward lower-energy candidates as virtual temperature decreases.
    }
    \label{fig_results}
\end{figure*}

In Fig.(\ref{fig_pipeline}b-d) we provide schematic diagrams depicting each of these key steps.
As shown in Fig.(\ref{fig_pipeline}b), the procedure begins with the identification of a reduced subspace based on conservation laws and symmetry properties.
Trial wavefunctions are then initialized, represented by the cross, triangle, and circle markers in Fig.(\ref{fig_pipeline}c).
Subsequently, the mixed-state ansatz undergoes iterative updates while the virtual temperature gradually decreases, filtering out higher-energy states and guiding the system toward the optimal solution. 
In Fig.(\ref{fig_pipeline}d), the trajectory of the markers represents the variational optimization process.
As $T$ decreases, higher-energy components (denoted by the blue cross and orange triangle) are progressively eliminated.
As depicted in Fig.(\ref{fig_pipeline}e), computational resources are dynamically reallocated to the most promising wavefunctions (indicated by the green circle), ensuring an efficient and effective search for the ground state.

The dynamic reallocation of measurement resources in KD-VQE is inspired by the softening mechanisms used in classical knowledge distillation (KD).
In classical KD, a teacher model generates a soft target distribution over classes, guiding the student model to focus on more informative outputs.
Similarly, in KD-VQE, a virtual temperature governs the distribution over trial wavefunctions, gradually allocating more measurements to lower-energy candidates.
The virtual temperature plays a crucial role in both frameworks.
In KD, temperature smooths the softmax output of the teacher model, providing richer learning signals.
In KD-VQE, virtual temperature determines the degree of mixing in the trial state ensemble, controlling the annealing schedule from a uniform mixture to a condensed state.


\section{Solving the 2-site Fermi-Hubbard Model}
To demonstrate the KD-VQE method in detail, we apply it to solving the ground state of the two-site Fermi-Hubbard model.

Hamiltonian of the two-site Fermi Hubbard model is given by\cite{hubbard1963electron}
\begin{equation}
    H_{hub} = 
    -t\sum_\sigma{\left(c_{1,\sigma}^\dagger c_{2, \sigma}+c_{2, \sigma}^\dagger c_{1, \sigma}\right)}
    + u\sum_{j =1,2}n_{j,\uparrow}n_{j,\downarrow}
    \label{eq_hubbard}
\end{equation}
where $t$ denotes the transfer integral, $u$ denotes the on-site interaction, and $\sigma =\uparrow,\downarrow$ indicates the spin.
For simplicity, here we set $\hbar=1$, and $t=1$, $u=1$.
This model maps naturally to four qubits, corresponding to spin-up and spin-down fermions on two lattice sites\cite{li2024iterative}.
In this study, we focus on the half-filling case, where the number of fermions is conserved.
Accordingly, the KD-VQE procedure targets subspaces containing two-particle states.

Following the identification of a reduced Hilbert subspace, the next step involves preparing the initial trial wavefunctions and constructing the corresponding mixed state.
Recalling Eq.(\ref{eq_hubbard}), the first term, which describes the hopping interaction, can be conveniently diagonalized by Fourier transformation\cite{verstraete2009quantum, cervera2018exact, li2023toward}.
Thus, we can use the eigenstates of the first term as initial guesses.
Fig.(\ref{fig_results}a) illustrates the implementation of the ansatz.
Each $U(\theta_k)$ denotes a set of single-qubit gates, parameterized by variational parameters $\theta_k$ for the $k^{th}$ trial wavefunction.
The $CNOT$ gates enclosed in the dashed box introduce entanglement and, together with the single-qubit gates, form a typical hardware-efficient ansatz\cite{kandala2017hardware, mitarai2019generalization, tilly2020computation, tilly2021reduced}.
The Fourier transformation is then applied, with its quantum circuit detailed in Fig.(\ref{fig_results}b).
We select six half-filling basis states as initial trial wavefunctions, labeled $\psi_{I,\cdots,VI}$.
Their configurations (prior to Fourier transformation) are illustrated in Fig.(\ref{fig_results}c).
These states can be easily initialized using the ansatz circuit from Fig.(\ref{fig_results}a).
For example, $\psi_I$, representing two fermions localized at site 1, is prepared by setting $U_{1,2}$ and $U_{3,2}$ to the Pauli-X gate and leaving all other single-qubit gates as identity operations.

During the optimization and virtual annealing process, there are in total $10^4$ shots.
Initially, the shots are uniformly distributed among the trial wavefunctions, corresponding to a maximally mixed state at $T\rightarrow+\infty$.
For each $\psi$, the gradient, along with the energy, are then estimated.
Variational parameters $\theta_k$ are updated per the gradient.
Here the simple gradient descent searching strategy\cite{tilly2022variational}, $\theta_k\rightarrow \theta_k -\eta\nabla\varepsilon_k$\cite{tilly2022variational}, where $\eta$ is the learning rate.
In general, KD-VQE benefits from a relatively large learning rate to accelerate the annealing process.
For the $k^{th}$ trial wavefunction, the optimization problem is given as
\begin{equation}
    \begin{aligned}
        \varepsilon_k =
    \mathop{\min}_{\theta_k}
    &\left\{
    \langle \psi_k({\theta_k})|H_{hub}|\psi_k({\theta_k})\rangle
    \right.
    \\
    &\left.-
    \lambda \left|\langle \psi_k({\theta_k})|\hat{n}|\psi_k({\theta_k})\rangle-2\right|
    \right\}
    \end{aligned}
\end{equation}
where $\lambda>0$ describes the penalty, and $\hat{n}$ is the particle number operator.
The penalty term\cite{kuroiwa2021penalty} is introduced to ensure that the optimization process concentrates on the half filling states.
Concurrently, the number of measurement shots assigned to each wavefunction is reallocated based on the Boltzmann weight $\exp(-\varepsilon_k/T)/Z$ , where $\varepsilon_k$ is the estimated energy of $\psi_k$.
The virtual temperature is initially set to $kT=25$, and is reduced by 5\% after each optimization step.
In Fig.(\ref{fig_results}d-g), we present the evolution of the estimated energy $\varepsilon$ during the optimization and virtual annealing process.
Trial wavefunctions receiving fewer than 100 shots are pruned from the pool.
As depicted in n Fig.(\ref{fig_results}d), $\psi_I$, which has the highest initial energy, is eliminated after approximately 60 steps.
$\psi_{II}$ and $\psi_{III}$exhibit similar energy trajectories and are filtered out around 80 steps.
The initial guesses of $\psi_{IV}$ and $\psi_{V}$ are already the local minima.
As shown in Fig.(\ref{fig_results}f), $\varepsilon_{IV}$ are $\varepsilon_V$ are trapped around 0 in the whole optimization process.
$\psi_{IV}$ and $\psi_{V}$ are removed around 85 steps.
In contrast, $\psi_{VI}$ is the most promising trial wavefunction.
By the end of KD-VQE, $\varepsilon_{VI}$ consistently demonstrates the lowest energy and converges toward the exact ground state energy of $-1.56$.
In Fig.(\ref{fig_results}h), we present the dynamic reallocation of measurement shots. The computational resources are progressively concentrated on the most promising candidate as virtual temperature $T$ decreases.
After around 85 steps, the measurement shots are all assigned to the most promising trial wavefunction $\psi_{VI}$, which corresponds to the blue bars in Fig.(\ref{fig_results}h).

In KD-VQE, measurement shots are distributed among multiple trial wavefunctions rather than concentrated on a single candidate. 
In general, estimating the energy within a specified error margin $\epsilon$ requires $\mathcal{O}(1/\epsilon^2)$ iterations and measurements.
Consequently, the energy estimate $\varepsilon_k$ becomes more accurate as more measurement shots are allocated to the corresponding trial wavefunction $\psi_k$.
As shown in Fig.(\ref{fig_results}g), $\varepsilon_{VI}$ becomes increasingly stable during the progression of the KD-VQE algorithm, as the measurement shots are reallocated toward $\psi_{VI}$.
In contrast, the estimates of other trial wavefunctions exhibit increasing fluctuations as the KD-VQE algorithm progresses, as shown in Fig.(\ref{fig_results}d-f).
By simultaneously evolving multiple trial wavefunctions, KD-VQE is capable to explore a broad variational landscape.

The virtual annealing schedule plays a critical role in determining the performance of KD-VQE.
Denote $s_c$ as the steps number of the optimization and virtual annealing, where subscript $c$ indicates that the mixed state condensates to the most promising state after $s_c$ steps.
In other words, the virtual temperature $T$ gradually decreases to zero within $s_c$ steps.
If the virtual temperature decreases too slowly, $s_c$ becomes excessively large, and suboptimal initial guesses may converge to local minima prior to the condensation.
In such cases, excessive computational resources may be expended on suboptimal candidates.
Conversely, if the virtual temperature is reduced too rapidly, resulting in a small $s_c$, the mixed state may undergo `premature condensation', potentially favoring a trial wavefunction that exhibits strong early performance but corresponds to a local rather than global minimum.

By the end of annealing, KD-VQE effectively filters out suboptimal states and identifies the most promising candidate.
This candidate can then be further refined through additional optimization to closely approximate the exact ground state.
In this context, KD-VQE remains compatible with most extensions and improvements developed within the standard VQE framework.

\section{Conclusions}
In this work, we introduce KD-VQE, a hybrid quantum-classical algorithm inspired by the virtual distillation process in KD.
KD-VQE enhances the conventional VQE framework by simultaneously evolving a set of trial wavefunctions.
Rather than relying on a single ansatz, KD-VQE maintains a mixture of candidate states, with measurement resources (shots) allocated according to a Boltzmann distribution governed by a virtual temperature.
As the temperature gradually decreases, resources are progressively concentrated on lower-energy candidates, thereby filtering out suboptimal states.
This virtual annealing process enables dynamic resource reallocation based on energy estimates, guiding the algorithm toward the global minimum.

To evaluate the performance of KD-VQE, we apply it to the two-site Fermi-Hubbard model.
Six representative trial wavefunctions are selected within the reduced subspace of half filling states.
A hardware-efficient ansatz is constructed for each wavefunction, and the KD-VQE is employed to refine the energy estimates and reallocate measurement resources.
The results show that KD-VQE effectively filters out high-energy candidates during the optimization process, ultimately converging to a trial state whose energy close to the exact ground state.

In conclusion, KD-VQE extends the standard VQE framework by integrating a virtual annealing process with variational optimization.
Through dynamic shots reallocation, KD-VQE improves convergence reliability, mitigates sensitivity to poor initializations, and explores a broader region of the solution space.
Our work sheds light on the promising synergy between knowledge distillation techniques and quantum variational methods, paving the way for more efficient and robust quantum optimization strategies.

{\it Acknowledgments} 
J.L gratefully acknowledges funding by National Natural Science Foundation of China (NSFC) under Grant No.12305012.

\bibliography{ref}

\begin{thebibliography}{45}%
\makeatletter
\providecommand \@ifxundefined [1]{%
 \@ifx{#1\undefined}
}%
\providecommand \@ifnum [1]{%
 \ifnum #1\expandafter \@firstoftwo
 \else \expandafter \@secondoftwo
 \fi
}%
\providecommand \@ifx [1]{%
 \ifx #1\expandafter \@firstoftwo
 \else \expandafter \@secondoftwo
 \fi
}%
\providecommand \natexlab [1]{#1}%
\providecommand \enquote  [1]{``#1''}%
\providecommand \bibnamefont  [1]{#1}%
\providecommand \bibfnamefont [1]{#1}%
\providecommand \citenamefont [1]{#1}%
\providecommand \href@noop [0]{\@secondoftwo}%
\providecommand \href [0]{\begingroup \@sanitize@url \@href}%
\providecommand \@href[1]{\@@startlink{#1}\@@href}%
\providecommand \@@href[1]{\endgroup#1\@@endlink}%
\providecommand \@sanitize@url [0]{\catcode `\\12\catcode `\$12\catcode `\&12\catcode `\#12\catcode `\^12\catcode `\_12\catcode `\%12\relax}%
\providecommand \@@startlink[1]{}%
\providecommand \@@endlink[0]{}%
\providecommand \url  [0]{\begingroup\@sanitize@url \@url }%
\providecommand \@url [1]{\endgroup\@href {#1}{\urlprefix }}%
\providecommand \urlprefix  [0]{URL }%
\providecommand \Eprint [0]{\href }%
\providecommand \doibase [0]{https://doi.org/}%
\providecommand \selectlanguage [0]{\@gobble}%
\providecommand \bibinfo  [0]{\@secondoftwo}%
\providecommand \bibfield  [0]{\@secondoftwo}%
\providecommand \translation [1]{[#1]}%
\providecommand \BibitemOpen [0]{}%
\providecommand \bibitemStop [0]{}%
\providecommand \bibitemNoStop [0]{.\EOS\space}%
\providecommand \EOS [0]{\spacefactor3000\relax}%
\providecommand \BibitemShut  [1]{\csname bibitem#1\endcsname}%
\let\auto@bib@innerbib\@empty
\bibitem [{\citenamefont {Peruzzo}\ \emph {et~al.}(2014)\citenamefont {Peruzzo}, \citenamefont {McClean}, \citenamefont {Shadbolt}, \citenamefont {Yung}, \citenamefont {Zhou}, \citenamefont {Love}, \citenamefont {Aspuru-Guzik},\ and\ \citenamefont {O’brien}}]{peruzzo2014variational}%
  \BibitemOpen
  \bibfield  {author} {\bibinfo {author} {\bibfnamefont {A.}~\bibnamefont {Peruzzo}}, \bibinfo {author} {\bibfnamefont {J.}~\bibnamefont {McClean}}, \bibinfo {author} {\bibfnamefont {P.}~\bibnamefont {Shadbolt}}, \bibinfo {author} {\bibfnamefont {M.-H.}\ \bibnamefont {Yung}}, \bibinfo {author} {\bibfnamefont {X.-Q.}\ \bibnamefont {Zhou}}, \bibinfo {author} {\bibfnamefont {P.~J.}\ \bibnamefont {Love}}, \bibinfo {author} {\bibfnamefont {A.}~\bibnamefont {Aspuru-Guzik}},\ and\ \bibinfo {author} {\bibfnamefont {J.~L.}\ \bibnamefont {O’brien}},\ }\bibfield  {title} {\bibinfo {title} {A variational eigenvalue solver on a photonic quantum processor},\ }\href@noop {} {\bibfield  {journal} {\bibinfo  {journal} {Nature communications}\ }\textbf {\bibinfo {volume} {5}},\ \bibinfo {pages} {4213} (\bibinfo {year} {2014})}\BibitemShut {NoStop}%
\bibitem [{\citenamefont {McClean}\ \emph {et~al.}(2016)\citenamefont {McClean}, \citenamefont {Romero}, \citenamefont {Babbush},\ and\ \citenamefont {Aspuru-Guzik}}]{mcclean2016theory}%
  \BibitemOpen
  \bibfield  {author} {\bibinfo {author} {\bibfnamefont {J.~R.}\ \bibnamefont {McClean}}, \bibinfo {author} {\bibfnamefont {J.}~\bibnamefont {Romero}}, \bibinfo {author} {\bibfnamefont {R.}~\bibnamefont {Babbush}},\ and\ \bibinfo {author} {\bibfnamefont {A.}~\bibnamefont {Aspuru-Guzik}},\ }\bibfield  {title} {\bibinfo {title} {The theory of variational hybrid quantum-classical algorithms},\ }\href@noop {} {\bibfield  {journal} {\bibinfo  {journal} {New Journal of Physics}\ }\textbf {\bibinfo {volume} {18}},\ \bibinfo {pages} {023023} (\bibinfo {year} {2016})}\BibitemShut {NoStop}%
\bibitem [{\citenamefont {Preskill}(2018)}]{preskill2018quantum}%
  \BibitemOpen
  \bibfield  {author} {\bibinfo {author} {\bibfnamefont {J.}~\bibnamefont {Preskill}},\ }\bibfield  {title} {\bibinfo {title} {Quantum computing in the nisq era and beyond},\ }\href@noop {} {\bibfield  {journal} {\bibinfo  {journal} {Quantum}\ }\textbf {\bibinfo {volume} {2}},\ \bibinfo {pages} {79} (\bibinfo {year} {2018})}\BibitemShut {NoStop}%
\bibitem [{\citenamefont {Bittel}\ and\ \citenamefont {Kliesch}(2021)}]{bittel2021training}%
  \BibitemOpen
  \bibfield  {author} {\bibinfo {author} {\bibfnamefont {L.}~\bibnamefont {Bittel}}\ and\ \bibinfo {author} {\bibfnamefont {M.}~\bibnamefont {Kliesch}},\ }\bibfield  {title} {\bibinfo {title} {Training variational quantum algorithms is np-hard},\ }\href@noop {} {\bibfield  {journal} {\bibinfo  {journal} {Physical review letters}\ }\textbf {\bibinfo {volume} {127}},\ \bibinfo {pages} {120502} (\bibinfo {year} {2021})}\BibitemShut {NoStop}%
\bibitem [{\citenamefont {O’Malley}\ \emph {et~al.}(2016)\citenamefont {O’Malley}, \citenamefont {Babbush}, \citenamefont {Kivlichan}, \citenamefont {Romero}, \citenamefont {McClean}, \citenamefont {Barends}, \citenamefont {Kelly}, \citenamefont {Roushan}, \citenamefont {Tranter}, \citenamefont {Ding} \emph {et~al.}}]{o2016scalable}%
  \BibitemOpen
  \bibfield  {author} {\bibinfo {author} {\bibfnamefont {P.~J.}\ \bibnamefont {O’Malley}}, \bibinfo {author} {\bibfnamefont {R.}~\bibnamefont {Babbush}}, \bibinfo {author} {\bibfnamefont {I.~D.}\ \bibnamefont {Kivlichan}}, \bibinfo {author} {\bibfnamefont {J.}~\bibnamefont {Romero}}, \bibinfo {author} {\bibfnamefont {J.~R.}\ \bibnamefont {McClean}}, \bibinfo {author} {\bibfnamefont {R.}~\bibnamefont {Barends}}, \bibinfo {author} {\bibfnamefont {J.}~\bibnamefont {Kelly}}, \bibinfo {author} {\bibfnamefont {P.}~\bibnamefont {Roushan}}, \bibinfo {author} {\bibfnamefont {A.}~\bibnamefont {Tranter}}, \bibinfo {author} {\bibfnamefont {N.}~\bibnamefont {Ding}}, \emph {et~al.},\ }\bibfield  {title} {\bibinfo {title} {Scalable quantum simulation of molecular energies},\ }\href@noop {} {\bibfield  {journal} {\bibinfo  {journal} {Physical Review X}\ }\textbf {\bibinfo {volume} {6}},\ \bibinfo {pages} {031007} (\bibinfo {year} {2016})}\BibitemShut {NoStop}%
\bibitem [{\citenamefont {Hempel}\ \emph {et~al.}(2018)\citenamefont {Hempel}, \citenamefont {Maier}, \citenamefont {Romero}, \citenamefont {McClean}, \citenamefont {Monz}, \citenamefont {Shen}, \citenamefont {Jurcevic}, \citenamefont {Lanyon}, \citenamefont {Love}, \citenamefont {Babbush} \emph {et~al.}}]{hempel2018quantum}%
  \BibitemOpen
  \bibfield  {author} {\bibinfo {author} {\bibfnamefont {C.}~\bibnamefont {Hempel}}, \bibinfo {author} {\bibfnamefont {C.}~\bibnamefont {Maier}}, \bibinfo {author} {\bibfnamefont {J.}~\bibnamefont {Romero}}, \bibinfo {author} {\bibfnamefont {J.}~\bibnamefont {McClean}}, \bibinfo {author} {\bibfnamefont {T.}~\bibnamefont {Monz}}, \bibinfo {author} {\bibfnamefont {H.}~\bibnamefont {Shen}}, \bibinfo {author} {\bibfnamefont {P.}~\bibnamefont {Jurcevic}}, \bibinfo {author} {\bibfnamefont {B.~P.}\ \bibnamefont {Lanyon}}, \bibinfo {author} {\bibfnamefont {P.}~\bibnamefont {Love}}, \bibinfo {author} {\bibfnamefont {R.}~\bibnamefont {Babbush}}, \emph {et~al.},\ }\bibfield  {title} {\bibinfo {title} {Quantum chemistry calculations on a trapped-ion quantum simulator},\ }\href@noop {} {\bibfield  {journal} {\bibinfo  {journal} {Physical Review X}\ }\textbf {\bibinfo {volume} {8}},\ \bibinfo {pages} {031022} (\bibinfo {year} {2018})}\BibitemShut {NoStop}%
\bibitem [{\citenamefont {Cao}\ \emph {et~al.}(2019)\citenamefont {Cao}, \citenamefont {Romero}, \citenamefont {Olson}, \citenamefont {Degroote}, \citenamefont {Johnson}, \citenamefont {Kieferov{\'a}}, \citenamefont {Kivlichan}, \citenamefont {Menke}, \citenamefont {Peropadre}, \citenamefont {Sawaya} \emph {et~al.}}]{cao2019quantum}%
  \BibitemOpen
  \bibfield  {author} {\bibinfo {author} {\bibfnamefont {Y.}~\bibnamefont {Cao}}, \bibinfo {author} {\bibfnamefont {J.}~\bibnamefont {Romero}}, \bibinfo {author} {\bibfnamefont {J.~P.}\ \bibnamefont {Olson}}, \bibinfo {author} {\bibfnamefont {M.}~\bibnamefont {Degroote}}, \bibinfo {author} {\bibfnamefont {P.~D.}\ \bibnamefont {Johnson}}, \bibinfo {author} {\bibfnamefont {M.}~\bibnamefont {Kieferov{\'a}}}, \bibinfo {author} {\bibfnamefont {I.~D.}\ \bibnamefont {Kivlichan}}, \bibinfo {author} {\bibfnamefont {T.}~\bibnamefont {Menke}}, \bibinfo {author} {\bibfnamefont {B.}~\bibnamefont {Peropadre}}, \bibinfo {author} {\bibfnamefont {N.~P.}\ \bibnamefont {Sawaya}}, \emph {et~al.},\ }\bibfield  {title} {\bibinfo {title} {Quantum chemistry in the age of quantum computing},\ }\href@noop {} {\bibfield  {journal} {\bibinfo  {journal} {Chemical reviews}\ }\textbf {\bibinfo {volume} {119}},\ \bibinfo {pages} {10856} (\bibinfo {year} {2019})}\BibitemShut {NoStop}%
\bibitem [{\citenamefont {Parrish}\ \emph {et~al.}(2019)\citenamefont {Parrish}, \citenamefont {Hohenstein}, \citenamefont {McMahon},\ and\ \citenamefont {Mart{\'\i}nez}}]{parrish2019quantum}%
  \BibitemOpen
  \bibfield  {author} {\bibinfo {author} {\bibfnamefont {R.~M.}\ \bibnamefont {Parrish}}, \bibinfo {author} {\bibfnamefont {E.~G.}\ \bibnamefont {Hohenstein}}, \bibinfo {author} {\bibfnamefont {P.~L.}\ \bibnamefont {McMahon}},\ and\ \bibinfo {author} {\bibfnamefont {T.~J.}\ \bibnamefont {Mart{\'\i}nez}},\ }\bibfield  {title} {\bibinfo {title} {Quantum computation of electronic transitions using a variational quantum eigensolver},\ }\href@noop {} {\bibfield  {journal} {\bibinfo  {journal} {Physical review letters}\ }\textbf {\bibinfo {volume} {122}},\ \bibinfo {pages} {230401} (\bibinfo {year} {2019})}\BibitemShut {NoStop}%
\bibitem [{\citenamefont {Quantum}\ \emph {et~al.}(2020)\citenamefont {Quantum}, \citenamefont {Collaborators*†}, \citenamefont {Arute}, \citenamefont {Arya}, \citenamefont {Babbush}, \citenamefont {Bacon}, \citenamefont {Bardin}, \citenamefont {Barends}, \citenamefont {Boixo}, \citenamefont {Broughton}, \citenamefont {Buckley} \emph {et~al.}}]{google2020hartree}%
  \BibitemOpen
  \bibfield  {author} {\bibinfo {author} {\bibfnamefont {G.~A.}\ \bibnamefont {Quantum}}, \bibinfo {author} {\bibnamefont {Collaborators*†}}, \bibinfo {author} {\bibfnamefont {F.}~\bibnamefont {Arute}}, \bibinfo {author} {\bibfnamefont {K.}~\bibnamefont {Arya}}, \bibinfo {author} {\bibfnamefont {R.}~\bibnamefont {Babbush}}, \bibinfo {author} {\bibfnamefont {D.}~\bibnamefont {Bacon}}, \bibinfo {author} {\bibfnamefont {J.~C.}\ \bibnamefont {Bardin}}, \bibinfo {author} {\bibfnamefont {R.}~\bibnamefont {Barends}}, \bibinfo {author} {\bibfnamefont {S.}~\bibnamefont {Boixo}}, \bibinfo {author} {\bibfnamefont {M.}~\bibnamefont {Broughton}}, \bibinfo {author} {\bibfnamefont {B.~B.}\ \bibnamefont {Buckley}}, \emph {et~al.},\ }\bibfield  {title} {\bibinfo {title} {Hartree-fock on a superconducting qubit quantum computer},\ }\href@noop {} {\bibfield  {journal} {\bibinfo  {journal} {Science}\ }\textbf {\bibinfo {volume} {369}},\ \bibinfo {pages} {1084} (\bibinfo {year} {2020})}\BibitemShut {NoStop}%
\bibitem [{\citenamefont {Cerezo}\ \emph {et~al.}(2021)\citenamefont {Cerezo}, \citenamefont {Arrasmith}, \citenamefont {Babbush}, \citenamefont {Benjamin}, \citenamefont {Endo}, \citenamefont {Fujii}, \citenamefont {McClean}, \citenamefont {Mitarai}, \citenamefont {Yuan}, \citenamefont {Cincio} \emph {et~al.}}]{cerezo2021variational}%
  \BibitemOpen
  \bibfield  {author} {\bibinfo {author} {\bibfnamefont {M.}~\bibnamefont {Cerezo}}, \bibinfo {author} {\bibfnamefont {A.}~\bibnamefont {Arrasmith}}, \bibinfo {author} {\bibfnamefont {R.}~\bibnamefont {Babbush}}, \bibinfo {author} {\bibfnamefont {S.~C.}\ \bibnamefont {Benjamin}}, \bibinfo {author} {\bibfnamefont {S.}~\bibnamefont {Endo}}, \bibinfo {author} {\bibfnamefont {K.}~\bibnamefont {Fujii}}, \bibinfo {author} {\bibfnamefont {J.~R.}\ \bibnamefont {McClean}}, \bibinfo {author} {\bibfnamefont {K.}~\bibnamefont {Mitarai}}, \bibinfo {author} {\bibfnamefont {X.}~\bibnamefont {Yuan}}, \bibinfo {author} {\bibfnamefont {L.}~\bibnamefont {Cincio}}, \emph {et~al.},\ }\bibfield  {title} {\bibinfo {title} {Variational quantum algorithms},\ }\href@noop {} {\bibfield  {journal} {\bibinfo  {journal} {Nature Reviews Physics}\ }\textbf {\bibinfo {volume} {3}},\ \bibinfo {pages} {625} (\bibinfo {year} {2021})}\BibitemShut {NoStop}%
\bibitem [{\citenamefont {Tilly}\ \emph {et~al.}(2022)\citenamefont {Tilly}, \citenamefont {Chen}, \citenamefont {Cao}, \citenamefont {Picozzi}, \citenamefont {Setia}, \citenamefont {Li}, \citenamefont {Grant}, \citenamefont {Wossnig}, \citenamefont {Rungger}, \citenamefont {Booth} \emph {et~al.}}]{tilly2022variational}%
  \BibitemOpen
  \bibfield  {author} {\bibinfo {author} {\bibfnamefont {J.}~\bibnamefont {Tilly}}, \bibinfo {author} {\bibfnamefont {H.}~\bibnamefont {Chen}}, \bibinfo {author} {\bibfnamefont {S.}~\bibnamefont {Cao}}, \bibinfo {author} {\bibfnamefont {D.}~\bibnamefont {Picozzi}}, \bibinfo {author} {\bibfnamefont {K.}~\bibnamefont {Setia}}, \bibinfo {author} {\bibfnamefont {Y.}~\bibnamefont {Li}}, \bibinfo {author} {\bibfnamefont {E.}~\bibnamefont {Grant}}, \bibinfo {author} {\bibfnamefont {L.}~\bibnamefont {Wossnig}}, \bibinfo {author} {\bibfnamefont {I.}~\bibnamefont {Rungger}}, \bibinfo {author} {\bibfnamefont {G.~H.}\ \bibnamefont {Booth}}, \emph {et~al.},\ }\bibfield  {title} {\bibinfo {title} {The variational quantum eigensolver: a review of methods and best practices},\ }\href@noop {} {\bibfield  {journal} {\bibinfo  {journal} {Physics Reports}\ }\textbf {\bibinfo {volume} {986}},\ \bibinfo {pages} {1} (\bibinfo {year} {2022})}\BibitemShut {NoStop}%
\bibitem [{\citenamefont {Sajjan}\ \emph {et~al.}(2022)\citenamefont {Sajjan}, \citenamefont {Li}, \citenamefont {Selvarajan}, \citenamefont {Sureshbabu}, \citenamefont {Kale}, \citenamefont {Gupta}, \citenamefont {Singh},\ and\ \citenamefont {Kais}}]{sajjan2022quantum}%
  \BibitemOpen
  \bibfield  {author} {\bibinfo {author} {\bibfnamefont {M.}~\bibnamefont {Sajjan}}, \bibinfo {author} {\bibfnamefont {J.}~\bibnamefont {Li}}, \bibinfo {author} {\bibfnamefont {R.}~\bibnamefont {Selvarajan}}, \bibinfo {author} {\bibfnamefont {S.~H.}\ \bibnamefont {Sureshbabu}}, \bibinfo {author} {\bibfnamefont {S.~S.}\ \bibnamefont {Kale}}, \bibinfo {author} {\bibfnamefont {R.}~\bibnamefont {Gupta}}, \bibinfo {author} {\bibfnamefont {V.}~\bibnamefont {Singh}},\ and\ \bibinfo {author} {\bibfnamefont {S.}~\bibnamefont {Kais}},\ }\bibfield  {title} {\bibinfo {title} {Quantum machine learning for chemistry and physics},\ }\href@noop {} {\bibfield  {journal} {\bibinfo  {journal} {Chemical Society Reviews}\ }\textbf {\bibinfo {volume} {51}},\ \bibinfo {pages} {6475} (\bibinfo {year} {2022})}\BibitemShut {NoStop}%
\bibitem [{\citenamefont {Ma}\ \emph {et~al.}(2023)\citenamefont {Ma}, \citenamefont {Liu}, \citenamefont {Shang}, \citenamefont {Fan}, \citenamefont {Li},\ and\ \citenamefont {Yang}}]{ma2023multiscale}%
  \BibitemOpen
  \bibfield  {author} {\bibinfo {author} {\bibfnamefont {H.}~\bibnamefont {Ma}}, \bibinfo {author} {\bibfnamefont {J.}~\bibnamefont {Liu}}, \bibinfo {author} {\bibfnamefont {H.}~\bibnamefont {Shang}}, \bibinfo {author} {\bibfnamefont {Y.}~\bibnamefont {Fan}}, \bibinfo {author} {\bibfnamefont {Z.}~\bibnamefont {Li}},\ and\ \bibinfo {author} {\bibfnamefont {J.}~\bibnamefont {Yang}},\ }\bibfield  {title} {\bibinfo {title} {Multiscale quantum algorithms for quantum chemistry},\ }\href@noop {} {\bibfield  {journal} {\bibinfo  {journal} {Chemical Science}\ }\textbf {\bibinfo {volume} {14}},\ \bibinfo {pages} {3190} (\bibinfo {year} {2023})}\BibitemShut {NoStop}%
\bibitem [{\citenamefont {Guo}\ \emph {et~al.}(2024)\citenamefont {Guo}, \citenamefont {Sun}, \citenamefont {Qian}, \citenamefont {Gong}, \citenamefont {Zhang}, \citenamefont {Chen}, \citenamefont {Ye}, \citenamefont {Wu}, \citenamefont {Cao}, \citenamefont {Liu} \emph {et~al.}}]{guo2024experimental}%
  \BibitemOpen
  \bibfield  {author} {\bibinfo {author} {\bibfnamefont {S.}~\bibnamefont {Guo}}, \bibinfo {author} {\bibfnamefont {J.}~\bibnamefont {Sun}}, \bibinfo {author} {\bibfnamefont {H.}~\bibnamefont {Qian}}, \bibinfo {author} {\bibfnamefont {M.}~\bibnamefont {Gong}}, \bibinfo {author} {\bibfnamefont {Y.}~\bibnamefont {Zhang}}, \bibinfo {author} {\bibfnamefont {F.}~\bibnamefont {Chen}}, \bibinfo {author} {\bibfnamefont {Y.}~\bibnamefont {Ye}}, \bibinfo {author} {\bibfnamefont {Y.}~\bibnamefont {Wu}}, \bibinfo {author} {\bibfnamefont {S.}~\bibnamefont {Cao}}, \bibinfo {author} {\bibfnamefont {K.}~\bibnamefont {Liu}}, \emph {et~al.},\ }\bibfield  {title} {\bibinfo {title} {Experimental quantum computational chemistry with optimized unitary coupled cluster ansatz},\ }\href@noop {} {\bibfield  {journal} {\bibinfo  {journal} {Nature Physics}\ }\textbf {\bibinfo {volume} {20}},\ \bibinfo {pages} {1240} (\bibinfo {year} {2024})}\BibitemShut {NoStop}%
\bibitem [{\citenamefont {Kim}\ \emph {et~al.}(2024)\citenamefont {Kim}, \citenamefont {Hu}, \citenamefont {Sohn}, \citenamefont {Kim}, \citenamefont {Kim}, \citenamefont {Lee},\ and\ \citenamefont {Lim}}]{kim2024qudit}%
  \BibitemOpen
  \bibfield  {author} {\bibinfo {author} {\bibfnamefont {B.}~\bibnamefont {Kim}}, \bibinfo {author} {\bibfnamefont {K.-M.}\ \bibnamefont {Hu}}, \bibinfo {author} {\bibfnamefont {M.-H.}\ \bibnamefont {Sohn}}, \bibinfo {author} {\bibfnamefont {Y.}~\bibnamefont {Kim}}, \bibinfo {author} {\bibfnamefont {Y.-S.}\ \bibnamefont {Kim}}, \bibinfo {author} {\bibfnamefont {S.-W.}\ \bibnamefont {Lee}},\ and\ \bibinfo {author} {\bibfnamefont {H.-T.}\ \bibnamefont {Lim}},\ }\bibfield  {title} {\bibinfo {title} {Qudit-based variational quantum eigensolver using photonic orbital angular momentum states},\ }\href@noop {} {\bibfield  {journal} {\bibinfo  {journal} {Science Advances}\ }\textbf {\bibinfo {volume} {10}},\ \bibinfo {pages} {eado3472} (\bibinfo {year} {2024})}\BibitemShut {NoStop}%
\bibitem [{\citenamefont {Delgado-Granados}\ \emph {et~al.}(2025)\citenamefont {Delgado-Granados}, \citenamefont {Krogmeier}, \citenamefont {Sager-Smith}, \citenamefont {Avdic}, \citenamefont {Hu}, \citenamefont {Sajjan}, \citenamefont {Abbasi}, \citenamefont {Smart}, \citenamefont {Narang}, \citenamefont {Kais} \emph {et~al.}}]{delgado2025quantum}%
  \BibitemOpen
  \bibfield  {author} {\bibinfo {author} {\bibfnamefont {L.~H.}\ \bibnamefont {Delgado-Granados}}, \bibinfo {author} {\bibfnamefont {T.~J.}\ \bibnamefont {Krogmeier}}, \bibinfo {author} {\bibfnamefont {L.~M.}\ \bibnamefont {Sager-Smith}}, \bibinfo {author} {\bibfnamefont {I.}~\bibnamefont {Avdic}}, \bibinfo {author} {\bibfnamefont {Z.}~\bibnamefont {Hu}}, \bibinfo {author} {\bibfnamefont {M.}~\bibnamefont {Sajjan}}, \bibinfo {author} {\bibfnamefont {M.}~\bibnamefont {Abbasi}}, \bibinfo {author} {\bibfnamefont {S.~E.}\ \bibnamefont {Smart}}, \bibinfo {author} {\bibfnamefont {P.}~\bibnamefont {Narang}}, \bibinfo {author} {\bibfnamefont {S.}~\bibnamefont {Kais}}, \emph {et~al.},\ }\bibfield  {title} {\bibinfo {title} {Quantum algorithms and applications for open quantum systems},\ }\href@noop {} {\bibfield  {journal} {\bibinfo  {journal} {Chemical Reviews}\ }\textbf {\bibinfo {volume} {125}},\ \bibinfo {pages} {1823} (\bibinfo {year} {2025})}\BibitemShut {NoStop}%
\bibitem [{\citenamefont {Kandala}\ \emph {et~al.}(2017)\citenamefont {Kandala}, \citenamefont {Mezzacapo}, \citenamefont {Temme}, \citenamefont {Takita}, \citenamefont {Brink}, \citenamefont {Chow},\ and\ \citenamefont {Gambetta}}]{kandala2017hardware}%
  \BibitemOpen
  \bibfield  {author} {\bibinfo {author} {\bibfnamefont {A.}~\bibnamefont {Kandala}}, \bibinfo {author} {\bibfnamefont {A.}~\bibnamefont {Mezzacapo}}, \bibinfo {author} {\bibfnamefont {K.}~\bibnamefont {Temme}}, \bibinfo {author} {\bibfnamefont {M.}~\bibnamefont {Takita}}, \bibinfo {author} {\bibfnamefont {M.}~\bibnamefont {Brink}}, \bibinfo {author} {\bibfnamefont {J.~M.}\ \bibnamefont {Chow}},\ and\ \bibinfo {author} {\bibfnamefont {J.~M.}\ \bibnamefont {Gambetta}},\ }\bibfield  {title} {\bibinfo {title} {Hardware-efficient variational quantum eigensolver for small molecules and quantum magnets},\ }\href@noop {} {\bibfield  {journal} {\bibinfo  {journal} {nature}\ }\textbf {\bibinfo {volume} {549}},\ \bibinfo {pages} {242} (\bibinfo {year} {2017})}\BibitemShut {NoStop}%
\bibitem [{\citenamefont {Kokail}\ \emph {et~al.}(2019)\citenamefont {Kokail}, \citenamefont {Maier}, \citenamefont {van Bijnen}, \citenamefont {Brydges}, \citenamefont {Joshi}, \citenamefont {Jurcevic}, \citenamefont {Muschik}, \citenamefont {Silvi}, \citenamefont {Blatt}, \citenamefont {Roos} \emph {et~al.}}]{kokail2019self}%
  \BibitemOpen
  \bibfield  {author} {\bibinfo {author} {\bibfnamefont {C.}~\bibnamefont {Kokail}}, \bibinfo {author} {\bibfnamefont {C.}~\bibnamefont {Maier}}, \bibinfo {author} {\bibfnamefont {R.}~\bibnamefont {van Bijnen}}, \bibinfo {author} {\bibfnamefont {T.}~\bibnamefont {Brydges}}, \bibinfo {author} {\bibfnamefont {M.~K.}\ \bibnamefont {Joshi}}, \bibinfo {author} {\bibfnamefont {P.}~\bibnamefont {Jurcevic}}, \bibinfo {author} {\bibfnamefont {C.~A.}\ \bibnamefont {Muschik}}, \bibinfo {author} {\bibfnamefont {P.}~\bibnamefont {Silvi}}, \bibinfo {author} {\bibfnamefont {R.}~\bibnamefont {Blatt}}, \bibinfo {author} {\bibfnamefont {C.~F.}\ \bibnamefont {Roos}}, \emph {et~al.},\ }\bibfield  {title} {\bibinfo {title} {Self-verifying variational quantum simulation of lattice models},\ }\href@noop {} {\bibfield  {journal} {\bibinfo  {journal} {Nature}\ }\textbf {\bibinfo {volume} {569}},\ \bibinfo {pages} {355} (\bibinfo {year} {2019})}\BibitemShut {NoStop}%
\bibitem [{\citenamefont {Uvarov}\ \emph {et~al.}(2020)\citenamefont {Uvarov}, \citenamefont {Biamonte},\ and\ \citenamefont {Yudin}}]{uvarov2020variational}%
  \BibitemOpen
  \bibfield  {author} {\bibinfo {author} {\bibfnamefont {A.}~\bibnamefont {Uvarov}}, \bibinfo {author} {\bibfnamefont {J.~D.}\ \bibnamefont {Biamonte}},\ and\ \bibinfo {author} {\bibfnamefont {D.}~\bibnamefont {Yudin}},\ }\bibfield  {title} {\bibinfo {title} {Variational quantum eigensolver for frustrated quantum systems},\ }\href@noop {} {\bibfield  {journal} {\bibinfo  {journal} {Physical Review B}\ }\textbf {\bibinfo {volume} {102}},\ \bibinfo {pages} {075104} (\bibinfo {year} {2020})}\BibitemShut {NoStop}%
\bibitem [{\citenamefont {Mizuta}\ \emph {et~al.}(2021)\citenamefont {Mizuta}, \citenamefont {Fujii}, \citenamefont {Fujii}, \citenamefont {Ichikawa}, \citenamefont {Imamura}, \citenamefont {Okuno},\ and\ \citenamefont {Nakagawa}}]{mizuta2021deep}%
  \BibitemOpen
  \bibfield  {author} {\bibinfo {author} {\bibfnamefont {K.}~\bibnamefont {Mizuta}}, \bibinfo {author} {\bibfnamefont {M.}~\bibnamefont {Fujii}}, \bibinfo {author} {\bibfnamefont {S.}~\bibnamefont {Fujii}}, \bibinfo {author} {\bibfnamefont {K.}~\bibnamefont {Ichikawa}}, \bibinfo {author} {\bibfnamefont {Y.}~\bibnamefont {Imamura}}, \bibinfo {author} {\bibfnamefont {Y.}~\bibnamefont {Okuno}},\ and\ \bibinfo {author} {\bibfnamefont {Y.~O.}\ \bibnamefont {Nakagawa}},\ }\bibfield  {title} {\bibinfo {title} {Deep variational quantum eigensolver for excited states and its application to quantum chemistry calculation of periodic materials},\ }\href@noop {} {\bibfield  {journal} {\bibinfo  {journal} {Physical Review Research}\ }\textbf {\bibinfo {volume} {3}},\ \bibinfo {pages} {043121} (\bibinfo {year} {2021})}\BibitemShut {NoStop}%
\bibitem [{\citenamefont {Gyawali}\ and\ \citenamefont {Lawler}(2022)}]{gyawali2022adaptive}%
  \BibitemOpen
  \bibfield  {author} {\bibinfo {author} {\bibfnamefont {G.}~\bibnamefont {Gyawali}}\ and\ \bibinfo {author} {\bibfnamefont {M.~J.}\ \bibnamefont {Lawler}},\ }\bibfield  {title} {\bibinfo {title} {Adaptive variational preparation of the fermi-hubbard eigenstates},\ }\href@noop {} {\bibfield  {journal} {\bibinfo  {journal} {Physical Review A}\ }\textbf {\bibinfo {volume} {105}},\ \bibinfo {pages} {012413} (\bibinfo {year} {2022})}\BibitemShut {NoStop}%
\bibitem [{\citenamefont {Anselme~Martin}\ \emph {et~al.}(2022)\citenamefont {Anselme~Martin}, \citenamefont {Simon},\ and\ \citenamefont {Ran{\v{c}}i{\'c}}}]{anselme2022simulating}%
  \BibitemOpen
  \bibfield  {author} {\bibinfo {author} {\bibfnamefont {B.}~\bibnamefont {Anselme~Martin}}, \bibinfo {author} {\bibfnamefont {P.}~\bibnamefont {Simon}},\ and\ \bibinfo {author} {\bibfnamefont {M.~J.}\ \bibnamefont {Ran{\v{c}}i{\'c}}},\ }\bibfield  {title} {\bibinfo {title} {Simulating strongly interacting hubbard chains with the variational hamiltonian ansatz on a quantum computer},\ }\href@noop {} {\bibfield  {journal} {\bibinfo  {journal} {Physical Review Research}\ }\textbf {\bibinfo {volume} {4}},\ \bibinfo {pages} {023190} (\bibinfo {year} {2022})}\BibitemShut {NoStop}%
\bibitem [{\citenamefont {Kattem{\"o}lle}\ and\ \citenamefont {Van~Wezel}(2022)}]{kattemolle2022variational}%
  \BibitemOpen
  \bibfield  {author} {\bibinfo {author} {\bibfnamefont {J.}~\bibnamefont {Kattem{\"o}lle}}\ and\ \bibinfo {author} {\bibfnamefont {J.}~\bibnamefont {Van~Wezel}},\ }\bibfield  {title} {\bibinfo {title} {Variational quantum eigensolver for the heisenberg antiferromagnet on the kagome lattice},\ }\href@noop {} {\bibfield  {journal} {\bibinfo  {journal} {Physical Review B}\ }\textbf {\bibinfo {volume} {106}},\ \bibinfo {pages} {214429} (\bibinfo {year} {2022})}\BibitemShut {NoStop}%
\bibitem [{\citenamefont {Wecker}\ \emph {et~al.}(2015)\citenamefont {Wecker}, \citenamefont {Hastings},\ and\ \citenamefont {Troyer}}]{wecker2015progress}%
  \BibitemOpen
  \bibfield  {author} {\bibinfo {author} {\bibfnamefont {D.}~\bibnamefont {Wecker}}, \bibinfo {author} {\bibfnamefont {M.~B.}\ \bibnamefont {Hastings}},\ and\ \bibinfo {author} {\bibfnamefont {M.}~\bibnamefont {Troyer}},\ }\bibfield  {title} {\bibinfo {title} {Progress towards practical quantum variational algorithms},\ }\href@noop {} {\bibfield  {journal} {\bibinfo  {journal} {Physical Review A}\ }\textbf {\bibinfo {volume} {92}},\ \bibinfo {pages} {042303} (\bibinfo {year} {2015})}\BibitemShut {NoStop}%
\bibitem [{\citenamefont {McClean}\ \emph {et~al.}(2018)\citenamefont {McClean}, \citenamefont {Boixo}, \citenamefont {Smelyanskiy}, \citenamefont {Babbush},\ and\ \citenamefont {Neven}}]{mcclean2018barren}%
  \BibitemOpen
  \bibfield  {author} {\bibinfo {author} {\bibfnamefont {J.~R.}\ \bibnamefont {McClean}}, \bibinfo {author} {\bibfnamefont {S.}~\bibnamefont {Boixo}}, \bibinfo {author} {\bibfnamefont {V.~N.}\ \bibnamefont {Smelyanskiy}}, \bibinfo {author} {\bibfnamefont {R.}~\bibnamefont {Babbush}},\ and\ \bibinfo {author} {\bibfnamefont {H.}~\bibnamefont {Neven}},\ }\bibfield  {title} {\bibinfo {title} {Barren plateaus in quantum neural network training landscapes},\ }\href@noop {} {\bibfield  {journal} {\bibinfo  {journal} {Nature communications}\ }\textbf {\bibinfo {volume} {9}},\ \bibinfo {pages} {4812} (\bibinfo {year} {2018})}\BibitemShut {NoStop}%
\bibitem [{\citenamefont {Wierichs}\ \emph {et~al.}(2020)\citenamefont {Wierichs}, \citenamefont {Gogolin},\ and\ \citenamefont {Kastoryano}}]{wierichs2020avoiding}%
  \BibitemOpen
  \bibfield  {author} {\bibinfo {author} {\bibfnamefont {D.}~\bibnamefont {Wierichs}}, \bibinfo {author} {\bibfnamefont {C.}~\bibnamefont {Gogolin}},\ and\ \bibinfo {author} {\bibfnamefont {M.}~\bibnamefont {Kastoryano}},\ }\bibfield  {title} {\bibinfo {title} {Avoiding local minima in variational quantum eigensolvers with the natural gradient optimizer},\ }\href@noop {} {\bibfield  {journal} {\bibinfo  {journal} {Physical Review Research}\ }\textbf {\bibinfo {volume} {2}},\ \bibinfo {pages} {043246} (\bibinfo {year} {2020})}\BibitemShut {NoStop}%
\bibitem [{\citenamefont {Hinton}\ \emph {et~al.}(2015)\citenamefont {Hinton}, \citenamefont {Vinyals},\ and\ \citenamefont {Dean}}]{hinton2015distilling}%
  \BibitemOpen
  \bibfield  {author} {\bibinfo {author} {\bibfnamefont {G.}~\bibnamefont {Hinton}}, \bibinfo {author} {\bibfnamefont {O.}~\bibnamefont {Vinyals}},\ and\ \bibinfo {author} {\bibfnamefont {J.}~\bibnamefont {Dean}},\ }\bibfield  {title} {\bibinfo {title} {Distilling the knowledge in a neural network},\ }\href@noop {} {\bibfield  {journal} {\bibinfo  {journal} {arXiv preprint arXiv:1503.02531}\ } (\bibinfo {year} {2015})}\BibitemShut {NoStop}%
\bibitem [{\citenamefont {Chen}\ \emph {et~al.}(2017)\citenamefont {Chen}, \citenamefont {Choi}, \citenamefont {Yu}, \citenamefont {Han},\ and\ \citenamefont {Chandraker}}]{chen2017learning}%
  \BibitemOpen
  \bibfield  {author} {\bibinfo {author} {\bibfnamefont {G.}~\bibnamefont {Chen}}, \bibinfo {author} {\bibfnamefont {W.}~\bibnamefont {Choi}}, \bibinfo {author} {\bibfnamefont {X.}~\bibnamefont {Yu}}, \bibinfo {author} {\bibfnamefont {T.}~\bibnamefont {Han}},\ and\ \bibinfo {author} {\bibfnamefont {M.}~\bibnamefont {Chandraker}},\ }\bibfield  {title} {\bibinfo {title} {Learning efficient object detection models with knowledge distillation},\ }\href@noop {} {\bibfield  {journal} {\bibinfo  {journal} {Advances in neural information processing systems}\ }\textbf {\bibinfo {volume} {30}} (\bibinfo {year} {2017})}\BibitemShut {NoStop}%
\bibitem [{\citenamefont {Cho}\ and\ \citenamefont {Hariharan}(2019)}]{cho2019efficacy}%
  \BibitemOpen
  \bibfield  {author} {\bibinfo {author} {\bibfnamefont {J.~H.}\ \bibnamefont {Cho}}\ and\ \bibinfo {author} {\bibfnamefont {B.}~\bibnamefont {Hariharan}},\ }\bibfield  {title} {\bibinfo {title} {On the efficacy of knowledge distillation},\ }in\ \href@noop {} {\emph {\bibinfo {booktitle} {Proceedings of the IEEE/CVF international conference on computer vision}}}\ (\bibinfo {year} {2019})\ pp.\ \bibinfo {pages} {4794--4802}\BibitemShut {NoStop}%
\bibitem [{\citenamefont {Cheng}\ \emph {et~al.}(2020)\citenamefont {Cheng}, \citenamefont {Rao}, \citenamefont {Chen},\ and\ \citenamefont {Zhang}}]{cheng2020explaining}%
  \BibitemOpen
  \bibfield  {author} {\bibinfo {author} {\bibfnamefont {X.}~\bibnamefont {Cheng}}, \bibinfo {author} {\bibfnamefont {Z.}~\bibnamefont {Rao}}, \bibinfo {author} {\bibfnamefont {Y.}~\bibnamefont {Chen}},\ and\ \bibinfo {author} {\bibfnamefont {Q.}~\bibnamefont {Zhang}},\ }\bibfield  {title} {\bibinfo {title} {Explaining knowledge distillation by quantifying the knowledge},\ }in\ \href@noop {} {\emph {\bibinfo {booktitle} {Proceedings of the IEEE/CVF conference on computer vision and pattern recognition}}}\ (\bibinfo {year} {2020})\ pp.\ \bibinfo {pages} {12925--12935}\BibitemShut {NoStop}%
\bibitem [{\citenamefont {Ji}\ and\ \citenamefont {Zhu}(2020)}]{ji2020knowledge}%
  \BibitemOpen
  \bibfield  {author} {\bibinfo {author} {\bibfnamefont {G.}~\bibnamefont {Ji}}\ and\ \bibinfo {author} {\bibfnamefont {Z.}~\bibnamefont {Zhu}},\ }\bibfield  {title} {\bibinfo {title} {Knowledge distillation in wide neural networks: Risk bound, data efficiency and imperfect teacher},\ }\href@noop {} {\bibfield  {journal} {\bibinfo  {journal} {Advances in Neural Information Processing Systems}\ }\textbf {\bibinfo {volume} {33}},\ \bibinfo {pages} {20823} (\bibinfo {year} {2020})}\BibitemShut {NoStop}%
\bibitem [{\citenamefont {Gu}\ \emph {et~al.}(2023)\citenamefont {Gu}, \citenamefont {Dong}, \citenamefont {Wei},\ and\ \citenamefont {Huang}}]{gu2023minillm}%
  \BibitemOpen
  \bibfield  {author} {\bibinfo {author} {\bibfnamefont {Y.}~\bibnamefont {Gu}}, \bibinfo {author} {\bibfnamefont {L.}~\bibnamefont {Dong}}, \bibinfo {author} {\bibfnamefont {F.}~\bibnamefont {Wei}},\ and\ \bibinfo {author} {\bibfnamefont {M.}~\bibnamefont {Huang}},\ }\bibfield  {title} {\bibinfo {title} {Minillm: Knowledge distillation of large language models},\ }\href@noop {} {\bibfield  {journal} {\bibinfo  {journal} {arXiv preprint arXiv:2306.08543}\ } (\bibinfo {year} {2023})}\BibitemShut {NoStop}%
\bibitem [{\citenamefont {Liu}\ \emph {et~al.}(2024)\citenamefont {Liu}, \citenamefont {Feng}, \citenamefont {Xue}, \citenamefont {Wang}, \citenamefont {Wu}, \citenamefont {Lu}, \citenamefont {Zhao}, \citenamefont {Deng}, \citenamefont {Zhang}, \citenamefont {Ruan} \emph {et~al.}}]{liu2024deepseek}%
  \BibitemOpen
  \bibfield  {author} {\bibinfo {author} {\bibfnamefont {A.}~\bibnamefont {Liu}}, \bibinfo {author} {\bibfnamefont {B.}~\bibnamefont {Feng}}, \bibinfo {author} {\bibfnamefont {B.}~\bibnamefont {Xue}}, \bibinfo {author} {\bibfnamefont {B.}~\bibnamefont {Wang}}, \bibinfo {author} {\bibfnamefont {B.}~\bibnamefont {Wu}}, \bibinfo {author} {\bibfnamefont {C.}~\bibnamefont {Lu}}, \bibinfo {author} {\bibfnamefont {C.}~\bibnamefont {Zhao}}, \bibinfo {author} {\bibfnamefont {C.}~\bibnamefont {Deng}}, \bibinfo {author} {\bibfnamefont {C.}~\bibnamefont {Zhang}}, \bibinfo {author} {\bibfnamefont {C.}~\bibnamefont {Ruan}}, \emph {et~al.},\ }\bibfield  {title} {\bibinfo {title} {Deepseek-v3 technical report},\ }\href@noop {} {\bibfield  {journal} {\bibinfo  {journal} {arXiv preprint arXiv:2412.19437}\ } (\bibinfo {year} {2024})}\BibitemShut {NoStop}%
\bibitem [{\citenamefont {Gou}\ \emph {et~al.}(2021)\citenamefont {Gou}, \citenamefont {Yu}, \citenamefont {Maybank},\ and\ \citenamefont {Tao}}]{gou2021knowledge}%
  \BibitemOpen
  \bibfield  {author} {\bibinfo {author} {\bibfnamefont {J.}~\bibnamefont {Gou}}, \bibinfo {author} {\bibfnamefont {B.}~\bibnamefont {Yu}}, \bibinfo {author} {\bibfnamefont {S.~J.}\ \bibnamefont {Maybank}},\ and\ \bibinfo {author} {\bibfnamefont {D.}~\bibnamefont {Tao}},\ }\bibfield  {title} {\bibinfo {title} {Knowledge distillation: A survey},\ }\href@noop {} {\bibfield  {journal} {\bibinfo  {journal} {International Journal of Computer Vision}\ }\textbf {\bibinfo {volume} {129}},\ \bibinfo {pages} {1789} (\bibinfo {year} {2021})}\BibitemShut {NoStop}%
\bibitem [{\citenamefont {Jattana}\ \emph {et~al.}(2023)\citenamefont {Jattana}, \citenamefont {Jin}, \citenamefont {De~Raedt},\ and\ \citenamefont {Michielsen}}]{jattana2023improved}%
  \BibitemOpen
  \bibfield  {author} {\bibinfo {author} {\bibfnamefont {M.~S.}\ \bibnamefont {Jattana}}, \bibinfo {author} {\bibfnamefont {F.}~\bibnamefont {Jin}}, \bibinfo {author} {\bibfnamefont {H.}~\bibnamefont {De~Raedt}},\ and\ \bibinfo {author} {\bibfnamefont {K.}~\bibnamefont {Michielsen}},\ }\bibfield  {title} {\bibinfo {title} {Improved variational quantum eigensolver via quasidynamical evolution},\ }\href@noop {} {\bibfield  {journal} {\bibinfo  {journal} {Physical Review Applied}\ }\textbf {\bibinfo {volume} {19}},\ \bibinfo {pages} {024047} (\bibinfo {year} {2023})}\BibitemShut {NoStop}%
\bibitem [{\citenamefont {Skogh}\ \emph {et~al.}(2023)\citenamefont {Skogh}, \citenamefont {Leinonen}, \citenamefont {Lolur},\ and\ \citenamefont {Rahm}}]{skogh2023accelerating}%
  \BibitemOpen
  \bibfield  {author} {\bibinfo {author} {\bibfnamefont {M.}~\bibnamefont {Skogh}}, \bibinfo {author} {\bibfnamefont {O.}~\bibnamefont {Leinonen}}, \bibinfo {author} {\bibfnamefont {P.}~\bibnamefont {Lolur}},\ and\ \bibinfo {author} {\bibfnamefont {M.}~\bibnamefont {Rahm}},\ }\bibfield  {title} {\bibinfo {title} {Accelerating variational quantum eigensolver convergence using parameter transfer},\ }\href@noop {} {\bibfield  {journal} {\bibinfo  {journal} {Electronic Structure}\ }\textbf {\bibinfo {volume} {5}},\ \bibinfo {pages} {035002} (\bibinfo {year} {2023})}\BibitemShut {NoStop}%
\bibitem [{\citenamefont {Hubbard}(1963)}]{hubbard1963electron}%
  \BibitemOpen
  \bibfield  {author} {\bibinfo {author} {\bibfnamefont {J.}~\bibnamefont {Hubbard}},\ }\bibfield  {title} {\bibinfo {title} {Electron correlations in narrow energy bands},\ }\href@noop {} {\bibfield  {journal} {\bibinfo  {journal} {Proceedings of the Royal Society of London. Series A. Mathematical and Physical Sciences}\ }\textbf {\bibinfo {volume} {276}},\ \bibinfo {pages} {238} (\bibinfo {year} {1963})}\BibitemShut {NoStop}%
\bibitem [{\citenamefont {Li}(2024)}]{li2024iterative}%
  \BibitemOpen
  \bibfield  {author} {\bibinfo {author} {\bibfnamefont {J.}~\bibnamefont {Li}},\ }\bibfield  {title} {\bibinfo {title} {Iterative method to improve the precision of the quantum-phase-estimation algorithm},\ }\href@noop {} {\bibfield  {journal} {\bibinfo  {journal} {Physical Review A}\ }\textbf {\bibinfo {volume} {109}},\ \bibinfo {pages} {032606} (\bibinfo {year} {2024})}\BibitemShut {NoStop}%
\bibitem [{\citenamefont {Verstraete}\ \emph {et~al.}(2009)\citenamefont {Verstraete}, \citenamefont {Cirac},\ and\ \citenamefont {Latorre}}]{verstraete2009quantum}%
  \BibitemOpen
  \bibfield  {author} {\bibinfo {author} {\bibfnamefont {F.}~\bibnamefont {Verstraete}}, \bibinfo {author} {\bibfnamefont {J.~I.}\ \bibnamefont {Cirac}},\ and\ \bibinfo {author} {\bibfnamefont {J.~I.}\ \bibnamefont {Latorre}},\ }\bibfield  {title} {\bibinfo {title} {Quantum circuits for strongly correlated quantum systems},\ }\href@noop {} {\bibfield  {journal} {\bibinfo  {journal} {Physical Review A—Atomic, Molecular, and Optical Physics}\ }\textbf {\bibinfo {volume} {79}},\ \bibinfo {pages} {032316} (\bibinfo {year} {2009})}\BibitemShut {NoStop}%
\bibitem [{\citenamefont {Cervera-Lierta}(2018)}]{cervera2018exact}%
  \BibitemOpen
  \bibfield  {author} {\bibinfo {author} {\bibfnamefont {A.}~\bibnamefont {Cervera-Lierta}},\ }\bibfield  {title} {\bibinfo {title} {Exact ising model simulation on a quantum computer},\ }\href@noop {} {\bibfield  {journal} {\bibinfo  {journal} {Quantum}\ }\textbf {\bibinfo {volume} {2}},\ \bibinfo {pages} {114} (\bibinfo {year} {2018})}\BibitemShut {NoStop}%
\bibitem [{\citenamefont {Li}\ \emph {et~al.}(2023)\citenamefont {Li}, \citenamefont {Jones},\ and\ \citenamefont {Kais}}]{li2023toward}%
  \BibitemOpen
  \bibfield  {author} {\bibinfo {author} {\bibfnamefont {J.}~\bibnamefont {Li}}, \bibinfo {author} {\bibfnamefont {B.~A.}\ \bibnamefont {Jones}},\ and\ \bibinfo {author} {\bibfnamefont {S.}~\bibnamefont {Kais}},\ }\bibfield  {title} {\bibinfo {title} {Toward perturbation theory methods on a quantum computer},\ }\href@noop {} {\bibfield  {journal} {\bibinfo  {journal} {Science Advances}\ }\textbf {\bibinfo {volume} {9}},\ \bibinfo {pages} {eadg4576} (\bibinfo {year} {2023})}\BibitemShut {NoStop}%
\bibitem [{\citenamefont {Mitarai}\ \emph {et~al.}(2019)\citenamefont {Mitarai}, \citenamefont {Yan},\ and\ \citenamefont {Fujii}}]{mitarai2019generalization}%
  \BibitemOpen
  \bibfield  {author} {\bibinfo {author} {\bibfnamefont {K.}~\bibnamefont {Mitarai}}, \bibinfo {author} {\bibfnamefont {T.}~\bibnamefont {Yan}},\ and\ \bibinfo {author} {\bibfnamefont {K.}~\bibnamefont {Fujii}},\ }\bibfield  {title} {\bibinfo {title} {Generalization of the output of a variational quantum eigensolver by parameter interpolation with a low-depth ansatz},\ }\href@noop {} {\bibfield  {journal} {\bibinfo  {journal} {Physical Review Applied}\ }\textbf {\bibinfo {volume} {11}},\ \bibinfo {pages} {044087} (\bibinfo {year} {2019})}\BibitemShut {NoStop}%
\bibitem [{\citenamefont {Tilly}\ \emph {et~al.}(2020)\citenamefont {Tilly}, \citenamefont {Jones}, \citenamefont {Chen}, \citenamefont {Wossnig},\ and\ \citenamefont {Grant}}]{tilly2020computation}%
  \BibitemOpen
  \bibfield  {author} {\bibinfo {author} {\bibfnamefont {J.}~\bibnamefont {Tilly}}, \bibinfo {author} {\bibfnamefont {G.}~\bibnamefont {Jones}}, \bibinfo {author} {\bibfnamefont {H.}~\bibnamefont {Chen}}, \bibinfo {author} {\bibfnamefont {L.}~\bibnamefont {Wossnig}},\ and\ \bibinfo {author} {\bibfnamefont {E.}~\bibnamefont {Grant}},\ }\bibfield  {title} {\bibinfo {title} {Computation of molecular excited states on ibm quantum computers using a discriminative variational quantum eigensolver},\ }\href@noop {} {\bibfield  {journal} {\bibinfo  {journal} {Physical Review A}\ }\textbf {\bibinfo {volume} {102}},\ \bibinfo {pages} {062425} (\bibinfo {year} {2020})}\BibitemShut {NoStop}%
\bibitem [{\citenamefont {Tilly}\ \emph {et~al.}(2021)\citenamefont {Tilly}, \citenamefont {Sriluckshmy}, \citenamefont {Patel}, \citenamefont {Fontana}, \citenamefont {Rungger}, \citenamefont {Grant}, \citenamefont {Anderson}, \citenamefont {Tennyson},\ and\ \citenamefont {Booth}}]{tilly2021reduced}%
  \BibitemOpen
  \bibfield  {author} {\bibinfo {author} {\bibfnamefont {J.}~\bibnamefont {Tilly}}, \bibinfo {author} {\bibfnamefont {P.}~\bibnamefont {Sriluckshmy}}, \bibinfo {author} {\bibfnamefont {A.}~\bibnamefont {Patel}}, \bibinfo {author} {\bibfnamefont {E.}~\bibnamefont {Fontana}}, \bibinfo {author} {\bibfnamefont {I.}~\bibnamefont {Rungger}}, \bibinfo {author} {\bibfnamefont {E.}~\bibnamefont {Grant}}, \bibinfo {author} {\bibfnamefont {R.}~\bibnamefont {Anderson}}, \bibinfo {author} {\bibfnamefont {J.}~\bibnamefont {Tennyson}},\ and\ \bibinfo {author} {\bibfnamefont {G.~H.}\ \bibnamefont {Booth}},\ }\bibfield  {title} {\bibinfo {title} {Reduced density matrix sampling: Self-consistent embedding and multiscale electronic structure on current generation quantum computers},\ }\href@noop {} {\bibfield  {journal} {\bibinfo  {journal} {Physical Review Research}\ }\textbf {\bibinfo {volume} {3}},\ \bibinfo {pages} {033230} (\bibinfo {year} {2021})}\BibitemShut {NoStop}%
\bibitem [{\citenamefont {Kuroiwa}\ and\ \citenamefont {Nakagawa}(2021)}]{kuroiwa2021penalty}%
  \BibitemOpen
  \bibfield  {author} {\bibinfo {author} {\bibfnamefont {K.}~\bibnamefont {Kuroiwa}}\ and\ \bibinfo {author} {\bibfnamefont {Y.~O.}\ \bibnamefont {Nakagawa}},\ }\bibfield  {title} {\bibinfo {title} {Penalty methods for a variational quantum eigensolver},\ }\href@noop {} {\bibfield  {journal} {\bibinfo  {journal} {Physical Review Research}\ }\textbf {\bibinfo {volume} {3}},\ \bibinfo {pages} {013197} (\bibinfo {year} {2021})}\BibitemShut {NoStop}%
\end{thebibliography}%

\end{document}